\documentstyle[preprint,aps,psfig]{revtex}
\begin{document}
\hsize\textwidth\columnwidth\hsize\csname@twocolumnfalse\endcsname

\draft

\title{Spin Relaxation of Conduction Electrons in Polyvalent Metals:
Theory and a Realistic Calculation}
\author{J. Fabian and S. Das Sarma}
\address{Department of Physics, University of Maryland at College Park,
College Park, MD 20742-4111}
\maketitle

\vspace{1cm}

\begin{abstract}

Relaxation of electronic spins in metals is significantly enhanced
whenever a Fermi surface crosses Brillouin zone boundaries, special
symmetry points, or lines of accidental degeneracy. A realistic calculation
shows that if aluminum had one valence electron, its spin relaxation
would be slower by nearly two orders of magnitude. This not only solves a
longstanding experimental puzzle, but also provides a way of
tailoring spin dynamics of electrons in a conduction band.

\end{abstract}
\vspace{2em}
\newpage

Electronic spin is emerging as a building block of new
digital devices. All-metal bipolar spin transistor\cite{johnson93a}
was already demonstrated and new devices are being built using the
phenomenon of giant magnetoresistance\cite{prinz95}.
Recent advances in spin-coherent dynamics in
semiconductors\cite{kikkawa97} may advance the development of a
quantum computer\cite{divincenzo95}. All these new technological
applications rely on relatively long relaxation time of
conduction electron spin eigenstates. In this Letter we consider
theoretically the problem of electronic spin relaxation in metals,
and solve an outstanding puzzle, explaining why very similar metals
(for example Al and Na) may have spin relaxation rates differing by
two or three orders of magnitude.

Experiments\cite{feher55,johnson85} show that spin
states in metals live several orders of magnitude longer than momentum
states. Furthermore, unlike momentum relaxation times $\tau$,
spin relaxation times $T_1$\cite{pines55} vary significantly among
metals. According to Elliott
\cite{elliott54} and Yafet \cite{yafet63} two
factors cause electronic spin in metals to decay:  (i) the
spin-orbit interaction induced, for example, by (crystal) ions or
impurities, and (ii) a momentum relaxation process such as impurity
or phonon scattering.  The product $c^2T_1$,
where $c^2$ is a measure of the strength of the spin-orbit
interaction, is then well approximated by $\tau$ and have similar
magnitude for all simple metals.
Considering only ion-induced spin-orbit interaction,
crystalline and atomic $c^2$ should be similar.
Indeed, by substituting the atomic values for $c^2$ Monod and
Beuneu \cite{monod79} found a ``Gr\"uneisen'' behavior of
$c^2T_1$ for all alkali and noble metals: as a function of
reduced temperature $T/T_D$, where $T_D$ is a Debye temperature, the
values of $c^2T_1$ fall onto a single curve.  On the other
hand, metals Al, Pd, Mg, and Be have $c^2T_1$ much smaller (up
to three orders of magnitude for Mg and Be) than the ``main group.''
For example, the
atomic $c^2$ for aluminum differs by less than
10\% from that for sodium \cite{monod79}, yet the spin relaxation
times at Debye temperatures 
are $0.1$ ns for aluminum
\cite{jaro1} ($T_D=394$ K \cite{ashcroft76}) and $20$ ns for sodium 
\cite{feher55,vescial64}($T_D=150$ K \cite{ashcroft76});
the corresponding momentum relaxation
times are in
the ratio 1:7\cite{allen87}.  We resolve this puzzle by showing that
the crystalline $c^2$ is about thirty times greater in Al than in
Na due to a rather subtle ``band renormalization'' effect.

In this paper we answer the question: ``Why spin in some metals
decays unexpectedly fast?'' by introducing the concept of {\it band
renormalized} spin-orbit interaction strength $c^2$.
Common to the ``strange'' metals (Al, Pd, Mg, and Be) with 
unusually fast spin relaxation  is that
their Fermi surfaces contain regions where $c^2$ is
significantly enhanced. Such regions are found near Brillouin zone
(BZ) boundaries, special symmetry points, or lines of accidental
degeneracy. Although these spin ``hot spots'' comprise,
generally, only a small part of the Fermi surface, they almost
entirely determine the effective value of $c^2$.
We predict that the ``strange'' behavior is,
in fact, common to all polyvalent metals, where spin hot spots are a
consequence of the Fermi surface topology. This prediction
is particularly significant since no polyvalent metals other than Al,
Pd, Mg, and Be have been measured for $T_1$ so far. Furthermore,
our calculations show that  $T_1$ can be appreciably altered by
modifying the band structure (by alloying, doping,
reducing the dimensionality, etc.). Our prediction is based on a
realistic pseudopotential calculation for aluminum and analytical 
estimates for $c^2$.

If the periodic potential due to ions in a crystal lattice
contains spin-orbit
coupling (a term proportional to the scalar product of the orbital
and spin momentum operators, $\hat{{\bf L}}\cdot\hat{{\bf S}}$),
the electronic Bloch states are a mixture of spin up
$\mid\uparrow\rangle$ and down $\mid\downarrow\rangle$
species\cite{elliott54}:
$\Psi^{\uparrow}_{{\bf k}n}({\bf r}) =[a_{{\bf
k}n}({\bf r})\mid\nolinebreak\uparrow\nolinebreak\rangle +b_{{\bf
k}n}({\bf r})\mid\nolinebreak\downarrow\nolinebreak
\rangle]\exp(i{\bf k}\cdot{\bf
r})
$ and
$\Psi^{\downarrow}_{{\bf k}n}({\bf r})=[a^*_{-{\bf k}n}({\bf
r})\mid\nolinebreak\downarrow\nolinebreak
\rangle -b^*_{-{\bf k}n}({\bf
r})\mid\nolinebreak\uparrow\nolinebreak\rangle]\exp(i{\bf k}\cdot{\bf
r}).  $
The lattice momentum ${\bf k}$ is confined to the first BZ,
$n$ is a band index, and $a_{{\bf k}n}({\bf r})$ and $b_{{\bf
k}n}({\bf r})$ are complex periodic functions with the period of the
lattice: if ${\bf G}$ denote the reciprocal lattice vectors, then
$a_{{\bf k}n}({\bf r})=\sum_{\bf G}a_{{\bf k}n}({\bf G})
\exp(i{\bf G}\cdot{\bf r})$ and
similarly for $b_{{\bf k}n}({\bf r})$. Both states have
the same energy $E_{{\bf k}n}$, as follows from time and space
inversion symmetry \cite{elliott54}; the numbering of bands is
therefore the same as without the spin notation.  The degenerate
states $\Psi^{\uparrow}_{{\bf k}n}$ and $\Psi^{\downarrow}_{{\bf
k}n}$ are chosen to represent electrons with spins polarized along
$z$ direction \cite{yafet63}:  $(\Psi^{\downarrow}_{{\bf
k}n}|\hat{S_z}| \Psi^{\downarrow}_{{\bf k}n})=
-(\Psi^{\uparrow}_{{\bf k}n}|\hat{S_z}| \Psi^{\uparrow}_{{\bf k}n})
< 0 $
and the off-diagonal matrix elements are zero. This condition implies
that $a_{{\bf k}n}({\bf r})$ have values close to one, while
$b_{{\bf k}n}({\bf r})$ are much smaller, decreasing with the decrease
of the strength of the spin-orbit interaction (with the exception of
the points where the spin-orbit interaction lifts a degeneracy).

Elliott\cite{elliott54} noticed that ordinary
(spin conserving) impurity
or phonon scattering can induce transitions between
$\Psi^{\uparrow}_{{\bf k}n}$ and $\Psi^{\downarrow}_{{\bf k'}n'}$,
leading to the flip of a spin polarization and thus spin relaxation.
If $\langle b^2 \rangle$ is the Fermi surface average of
$|b_{{\bf k}n}|^2=\sum_{{\bf G}}|b_{{\bf k}n}({\bf G})|^2$,
spin relaxation rate $1/T_1\approx 4\langle b^2 \rangle 1/\tau$
(that is, $c^2\approx 4\langle b^2 \rangle$). Such
a formula could be obtained by assuming that (1)
$1/\tau \gg 1/T_1$ \cite{overhauser53}, (2) $a_{{\bf k}n}({\bf
r})\approx 1$, (3) scattering form factor has a constant
amplitude, and (4) the
interference between $b_{{\bf k}n}({\bf G})$ with different ${\bf G}$
is neglected. Assumptions (1) and (2) are usually satisfied.
Assumption (3) implies a scattering by a delta-function-like impurity
potential. This simplification is consistent with
our goal to establish the effect of band-structure
(through $b_{{\bf k}n}$) on $T_1$ rather than to study
particular scattering processes.
The neglect of interference can be justified
if the scattering form factors have rapidly varying phase.
Although such a form factor [that also satisfies (3)] is hardly
found, realistic form factors do oscillate on the momentum
scale where the transitions occur. Assumption (4) is then
partly justified, since the phase attached to $b_{{\bf k}n}({\bf
G})$ will be different for different ${\bf G}$.

The spin-mixing parameters $|b_{{\bf k}n}|^2$ can have a broad
range of values, depending on the position of ${\bf k}$ in
the BZ. Consider a band structure computed
without the spin-orbit interaction. If the closest band to 
$n$ is separated from $n$ by $\Delta$, 
the spin-orbit interaction mixes the spins from the
two bands (direct interband transitions): $|b_{{\bf k}n}|^2 \approx
(1-\Delta/\sqrt{\Delta^2+4V^2_{SO}})/2$, where
$V_{SO}$ is some effective spin-orbit interaction.
Three cases occur. (A) For a general point, the band
separation is of order $E_F$, the Fermi energy, so that $\Delta\gg
V_{SO}$ and $|b_{{\bf k}n}|^2\approx (V_{SO}/E_F)^2$.
(B) If the state is close to a BZ plane that cuts ${\bf G}$ by
half, the band separation is $\approx 2V_{\bf G}$ ($V_{\bf G}$ is the
${\bf G}${\it th} Fourier coefficient of the non-spin part of the
lattice potential). Since typically $V_{\bf G}\gg V_{SO}$,
$|b_{{\bf k}n}|^2\approx (V_{SO}/2V_{\bf G})^2$; this can be a few
orders larger than in (A).  Finally, (C) the spin-orbit interaction
can lift the degeneracy of two or more bands. 
The mixing of spins is complete and $|b_{{\bf k}n}|^2\approx |a_{{\bf
k}n}|^2\approx 0.5$. The states with property (C) are not covered
by the Elliott-Yafet theory\cite{yafet63}. This is not a concern,
however, since as we show below, such states are statistically
irrelevant for aluminum $T_1$.  

To illustrate how the band structure affects $\langle b^2 \rangle$ 
(and $T_1$), we perform a pseudopotential calculation for aluminum, 
where all  three cases (A) to (C) occur. Our pseudopotential has a 
non-spin (scalar) part and a spin-orbit term.  The scalar 
part\cite{masovic78} has the nice feature [in view of estimate (B)] 
of being fitted to the experimental values \cite{ashcroft63} of 
$V_1=V_{111}=0.00895$ a.\,u.
and $V_2=V_{200}=0.0281$ a.\,u.  ($1$ a.\,u. = $2$ Ry). In addition to
accurately reproducing the band structure of aluminum this form
factor gives reasonable results\cite{masovic78} for the resistivity of liquid
aluminum, making it useful in scattering problems.
The spin-orbit part of our pseudopotential is
$\lambda \hat{\bf L}\cdot\hat{\bf S}{\cal P}_1$,
where ${\cal P}_l$ is the operator projecting on the orbital
momentum state $l$. The parameter $\lambda=2.7\times 10^{-3}$ a.\,u.
inside the ion core of twice the Bohr radius, $r_c=2r_B$. Outside the
core $\lambda$ vanishes. This estimate for $\lambda$ is based on the
first-principles spin-orbit pseudopotential of Ref. \cite{bachelet82}.
Since the spin-orbit interaction acts only inside the core,
the effective $V_{SO}$ is reduced to about  $(r^3_c/V_a)\lambda$
($\approx \lambda/10$ for aluminum, where the primitive cell volume
$V_a\approx 111 r^3_B$).

Having the band structure,
the challenge is to evaluate the average of $|b_{{\bf k}n}|^2$,
a quantity that varies over several orders of magnitudes.
We do the averaging by the tetrahedron method\cite{allen83}
 with a carefully
designed grid that envelops the two sheets of the aluminum
Fermi surface (bands 2 and 3). The grid is denser in the regions
where $\Delta$ is smaller (around BZ boundaries and accidental
degeneracy points). This adaptation of the grid is necessary
to ensure that $|b_{{\bf k}n}|^2$ can be linearly interpolated inside
the grid cells (tetrahedrons), as assumed in the tetrahedron method.
The number of grid points (in the irreducible wedge of the
first BZ) where the band-structure equations must be solved is
about ten thousand.

Figure \ref{fig:1} plots the calculated distribution
$\rho$ of the values attainable by $|b_{{\bf k}n}|^2$ over
the Fermi  surface 
(the visual perspective of the distribution is in Fig. 
\ref{fig:2}).
The span is enormous--almost seven decades!
The majority of states have $|b_{{\bf k}n}|^2$ below $10^{-5}$. These
are the generic points from the estimate (A). Once the Fermi surface
approaches the BZ planes (violet in Fig. \ref{fig:2}), 
the values jump to $10^{-5}-10^{-4}$. However, the largest
$|b_{{\bf k}n}|^2$ are found near  the
accidental degeneracy points R (red spots in Fig. \ref{fig:2});
these points do not lie on symmetry lines, yet they
have degenerate bands\cite{harrison66} if no spin-orbit interaction
is present. The spin-orbit interaction lifts this degeneracy and
completely mixes the spin states as in the estimate (C).
The unusually long tail of $\rho$ ensures the
large value of the average $\langle b^2 \rangle \approx2.0\times
10^{-5}$, about ten times larger than the value where $\rho$
is maximal. To ensure that the above picture is valid for a large
temperature range, we checked that spin hot spots survive energy
excitations of at least $\approx kT_D$ ($\approx 35$ meV) above 
and below the Fermi surface. This also invalidates the objection
that small relativistic corrections (of a few meV) could affect 
the existence of the spin hot spots in aluminum.

The band structure role in $\langle b^2 \rangle$
becomes even more evident from the plot (Fig. \ref{fig:1})
of $\rho$ for a hypothetical case of
monovalent aluminum (the lattice and the form factors are unchanged).
The monovalent aluminum has a simple Fermi
surface, without deformations of type (B) or (C).
The distribution of $b^2$ is appropriately narrower, and the average
value $\langle b^2 \rangle\approx 3.4\times 10^{-7}$ is about {\it
fifty} times smaller than for trivalent aluminum. Adjusting for the
density of states (monovalent aluminum would have 
$1/\tau$ reduced $\sim 1/3^{1/3}$ times) the spin relaxation would be 
about seventy times slower. These values may somewhat vary for 
different scattering processes.

How different spin ``hot spots'' contribute to 
the renormalization of $\langle b^2 \rangle$ becomes clear from
Fig. \ref{fig:3}, which plots what we call 
the {\it average density of bands} (ADOB). ADOB is the Fermi surface
average of the number of (vertical) bands in the interval 
$(\Delta, \Delta + d\Delta)$: 
$ADOB(\Delta)=(1/g_F)\sum_{{\bf k},n\ne m}\delta({\tilde E}_{{\bf
k}n}-E_F) \delta(|{\tilde E}_{{\bf k}m}-E_F
|-\Delta)$, were band energies ${\tilde E}_{{\bf k}n}$
are computed without the spin-orbit interaction and  
$g_F$ is the density of states (per spin) at $E_F$.
The average $\langle b^2 \rangle$ can then be well approximated
by taking the integral (from zero to infinity) of ADOB weighted 
by the mixing factor $(1-\Delta/\sqrt{\Delta^2+4V^2_{SO}})/2$.
Figure \ref{fig:3} uses $V_{SO}=
1.1\times 10^{-4}$ a.\,u., the value that gives the right answer
for $\langle b^2 \rangle$.
At small $\Delta$ ADOB is linear. This is expected for a 
region around R where band gaps increase linearly with 
increasing distance from R \cite{harrison66}. 
The linear increase (see also Tab. \ref{tab:1})
continues up to the point where the Fermi surface crosses the first
($\Delta\approx 2V_{1}$) and the second ($\Delta\approx2V_{2}$)
closest BZ boundary plane. At these points ADOB has  
power-law singularities (Tab. \ref{tab:1}).
Larger band separations, where ADOB
develops either logarithmic or step-like singularities\cite{jaro2},
are  irrelevant. Indeed, the cumulative average $\langle b^2
\rangle(\Delta)$ saturates after the second peak so that $\langle
b^2 \rangle$ is almost entirely determined by the regions close
to accidental degeneracies and BZ boundaries. As Fig. \ref{fig:3}
shows, these regions contribute about equally to $\langle b^2 \rangle$.
The behavior of $\langle b^2\rangle(\Delta)$ also shows that states
in the immediate neighborhood of R (the red spots in Fig.\ref{fig:2})
with $\Delta \alt 2V_{SO}\approx 2.2\times 10^{-4}$ are statistically
irrelevant. It is rather a broader neighborhood of R [the states
with $2V_{SO} \alt \Delta \alt 2V_1$ and estimate (B)] that are 
contributing most. These findings are confirmed by the analytical 
calculation reported in Tab. \ref{tab:1}.

Table \ref{tab:1} summarizes our estimates of ADOB and
$\langle b^2 \rangle$ for three different cases. The estimates were
obtained analytically by the orthogonalized-plane-wave
methods developed in Ref. \cite{harrison66}. If the Fermi surface
(in an extended zone) 
lies entirely within the first BZ (region I, alkali and noble metals),
the enhancement of $\langle b^2 \rangle$ is about tenfold
since $E_{MIN}\approx 0.1$.
If the Fermi surface crosses a zone boundary (region
II, polyvalent metals) the enhancement is $\sim 1/V_{\bf G}$, typically a
hundred. Relative to I, however, the enhancement is in the units of
ten, in agreement with the numerical calculation.
Qualitatively, a fraction of $\sim V_{\bf G}$ states on the Fermi surface
comes close to zone boundaries ($\Delta\approx 2V_{\bf G}$) so that their
$|b_{{\bf k}n}|^2\sim(V_{SO}/V_{\bf G})^2$. The contribution
of these points to $\langle b^2 \rangle$ is therefore $\sim 
V_{\bf G}\times (V_{SO}/V_{\bf G})^2$; the enhancement is 
$\sim 1/V_{\bf G}$, consistent with the result
in Tab. \ref{tab:1}.
Curiously, the region II applies also to the noble metals,
whose Fermi surfaces touch zone boundaries\cite{ashcroft76}.
This effect is, however, masked by the unusually high $V_{\bf G}$ for
these metals \cite{cohen70}
($V_{\bf G}\approx E_{MIN}$, otherwise the Fermi surface would not touch
the zone planes).
The enhancement of $\langle b^2 \rangle$ can be also significant
if the Fermi surface crosses a line
of accidental degeneracy (region III). Indeed, by substituting
the values for aluminum, the enhancement is about tenfold,
similar to case (II), again agreeing with the numerical result. 
Finally, not shown in the Tab. \ref{tab:1} is the case when $E_F$ 
coincides with a degenerate level at a special symmetry point. Such a 
situation occurs, for example, in Pd and Pt, whose Fermi surfaces
go through the fcc L point.
If the spin-orbit interaction lifts this
degeneracy, the renormalization of $\langle b^2 \rangle$ can
be significant (we find the enhancement $\sim V_{\bf G}/V_{SO}$
for the fcc W point\cite{jaro2}).

\begin{table}
\begin{tabular}{|lccr|}
FS & multiplicative & ADOB & $b^2/V^2_{SO}$ \\
area &       factor &  &  \\ \hline
I. & $ N_G/4Gk_F$&$ \Theta(\Delta-E_{MIN})$ &
$1/E_{MIN}$ \\
II. & $N_G/4Gk_F$&$\Delta/\sqrt{\Delta^2-4V_{\bf G}^2}$ &
$(\pi/4)(1/V_{\bf G})$ \\ III. &
$N_R/8k_F$&$\left(V_{2}/V_{1}\right)^2\Delta$ &
$\left(V_2/V_1\right)^2
\ln (V_1/V_{SO})$
\end{tabular}
\caption{Estimated contributions of different
Fermi surface  regions to ADOB and $\langle b^2 \rangle$.
(I) The Fermi surface is
entirely within the first BZ, crosses (II) a zone
boundary, and (III) an accidental degeneracy line at R.
Momentum and energy is measured in the units of $G/2$ and
$(G/2)^2/2$, respectively. The region III assumes an
fcc lattice with period $a$, energy and momentum in the
units of $2\pi/a$ and $(2\pi/a)^2/2$, respectively, and 
$V_2 \gg V_1 \gg V^2_2$.
Both ADOB and $\langle b^2 \rangle$ come with the corresponding
multiplicative factor.
Notation: $k_F=\sqrt{E_F}$ is the Fermi vector,
$E_{MIN}=G(G-2k_F)$ is the band gap at the point of closest
approach to the BZ plane given by ${\bf G}$;
$N_G$, and $N_R$ are the
numbers of the corresponding BZ planes and accidental
degeneracy points $(N_R=24)$;
$\Theta$ is the step function.}
\label{tab:1}
\end{table}

Our final note concerns the hexagonal Mg and Be, where the
deviation of $T_1$ from the ``main group'' is most striking
\cite{monod79}. We argue that this is also a manifestation
of the band renormalization of $c^2$. Without the spin-orbit
interaction, all the states at the hexagonal faces of the first 
BZ
of a simple hexagonal structure are degenerate \cite{ashcroft76}.
The spin-orbit interaction lifts this degeneracy \cite{elliott54} (except
at some symmetry points and lines),
presumably by the amount $V_{\bf G}V_{SO}$, the
largest second-order term containing the spin-orbit interaction
(any first order term vanishes since the structure factor
associated with the hexagonal faces is zero\cite{ashcroft76}).
The contribution to $\langle b^2 \rangle$ of the points where
the Fermi surface intersects the hexagonal faces is 
$\sim V_{\bf G}V_{SO}$ (in the units of Tab. \ref{tab:1}):
the characteristic value $|b_{{\bf k}n}|^2
\sim 1$, times the area of the affected part of the Fermi surface,
$V_{\bf G}V_{SO}$. The enhancement measured in terms
of $V^2_{SO}$ is then $V_{\bf G}/V_{SO}$; this can be as large as
a thousand for light elements like Mg and Be.

We acknowledge discussions with P. B. Allen and M. Johnson.      
This work was supported by the U.S. ONR.

{\it Note Added.}--After submission of our work R. H. Silsbee brought
to our attention an earlier paper (R. H. Silsbee and F. Beuneu,
Phys. Rev. B {\bf 27}, 2682, 1983) which suggests the
importance of the  accidental degeneracy points for
the spin relaxation in aluminum.

\begin{figure}
\caption{Calculated distribution $\rho$ (in arbitrary units)
of the spin-mixing parameters
$|b_{{\bf k}n}|^2$ for aluminum. The corresponding average
$\langle b^2\rangle\approx2.0\times10^{-5}$ is indicated by solid arrow.
The linear tail of the distribution is shown in the inset.
The dashed line shows what the distribution would
be if aluminum were monovalent ($\langle b^2\rangle\approx3.4\times10^{-7}$,
dashed arrow).
}
\label{fig:1}
\end{figure}

\begin{figure}
\caption{Stereogram of the Fermi momentum directions in aluminum.
The fragment shows the spin ``hot spots'': the points ${\bf k}$
(in an extended-zone scheme) with $|b_{{\bf k}n}|^2
\ge 10^{-5}$.
Colors violet, blue, green, yellow,
and red indicate a successive increase of $b^2$:
violet points have $b^2$ from $10^{-5}$ to $10^{-4}$, blue
$10^{-4}$ to $10^{-3}$, etc., up to $10^{-1}-1$ for red.
To improve their visibility, the weight of the colors (except
for violet) is enhanced.
}
\label{fig:2}
\end{figure}

\begin{figure}
\caption{Calculated average density of bands (ADOB)
and cumulative average spin-mixing parameter $\langle b^2\rangle(
\Delta)$
for aluminum.
}
\label{fig:3}
\end{figure}


\begin{references}
\bibitem{johnson93a} M. Johnson, Science {\bf 260}, 320 (1993);
J. Magn. Magn. Mater. {\bf 140-144}, 21 (1995); {\it Ibid.} 156 (1996);
 Mater. Sci. Eng.  B {\bf 31}, 199 (1995).
\bibitem{prinz95} G. Prinz, Phys. Today {\bf 48}, 58 (1995).
\bibitem{kikkawa97} J. M. Kikkawa, I. P. Smorchkova, N. Samarth,
and D. D. Awschalom, Science {\bf 277}, 1284 (1997); J. M. Kikkawa
and D. D. Awschalom, Phys. Rev. Lett. {\bf 80}, 4313 (1998).
\bibitem{divincenzo95} D. P. DiVincenzo, Science {\bf 270}, 255 (1995).
\bibitem{feher55} G. Feher and A. F. Kip,
Phys. Rev. {\bf 98}, 337 (1955).
\bibitem{johnson85} M. Johnson and R. H. Silsbee, Phys. Rev. Lett.
{\bf 55}, 1790 (1985); Phys. Rev. B {\bf 37}, 5326 (1988).
\bibitem{pines55} The longitudinal ``spin-lattice'' relaxation time
$T_1$ equals, at least for cubic metals, the transverse ``dephasing''
relaxation time $T_2$, as shown in D. Pines and C. P. Slichter,
Phys. Rev. {\bf 100},
1014 (1955).
\bibitem{elliott54} R. J. Elliott, Phys. Rev. {\bf 96},
266 (1954).
\bibitem{yafet63} Y. Yafet, in {\it Solid State Physics}, edited by F.
Seitz and D. Turnbull (Academic, New York, 1963), Vol. 14.
\bibitem{monod79} P. Monod and F. Beuneu, Phys. Rev. B {\bf 19},
911 (1979); F. Beuneu and P. Monod, Phys. Rev. B {\bf 18},
2422 (1978).
\bibitem{jaro1} Extrapolated from the experimental data in Ref.
\cite{johnson85} and S. Schultz, G. Dunifer, and C. Latham, Phys. Lett.
{\bf 23}, 192 (1966); D. Lubzens and S. Schultz, Phys. Rev. Lett. {\bf 36},
1104 (1976).
\bibitem{ashcroft76} N. W. Ashcroft and N. D. Mermin, {\it Solid
State Physics}, (Saunders, New York, 1976).
\bibitem{vescial64} F. Vescial, N. S. Vander Ven, and R. T. Schumacher,
Phys. Rev. {\bf 134}, A1286 (1964); W. Kolbe, Phys. Rev. B {\bf 3},
320 (1971).
\bibitem{allen87} P. B. Allen, Phys. Rev. B {\bf 36},
2920 (1987).
\bibitem{overhauser53} A. W. Overhauser, Phys. Rev. {\bf 89},
689 (1953).
\bibitem{masovic78} D. R. Ma\v{s}ovi\'{c} and S. Zekovi\'{c},
phys. stat. sol. (b)
{\bf 89}, K57 (1978); V. Veljkovi\'{c} and I. Slavi\'{c},
Phys. Rev. Lett. {\bf 29}, 105 (1972).
\bibitem{ashcroft63} N. W. Ashcroft, Phil. Mag. {\bf 8}, 2055 (1963).
\bibitem{bachelet82} G. B. Bachelet, D. R. Hamann, and M. Schl\"{u}ter,
Phys. Rev. B {\bf 26}, 4199 (1982).
\bibitem{allen83} P. B. Allen, phys. stat. sol. (b) {\bf 120}, 529 (1983).
\bibitem{harrison66} W. A. Harrison, {\it Pseudopotentials in the Theory
of Metals}, (Benjamin, New York, 1966).
\bibitem{jaro2} J. Fabian and S. Das Sarma (unpublished).
\bibitem{cohen70} M. L. Cohen and V. Heine, in {\it Solid State Physics}, 
edited by H. Ehrenreich, F. Seitz, and
D. Turnbull (Academic, New York, 1970), Vol. 24, p. 183.
\end{references}
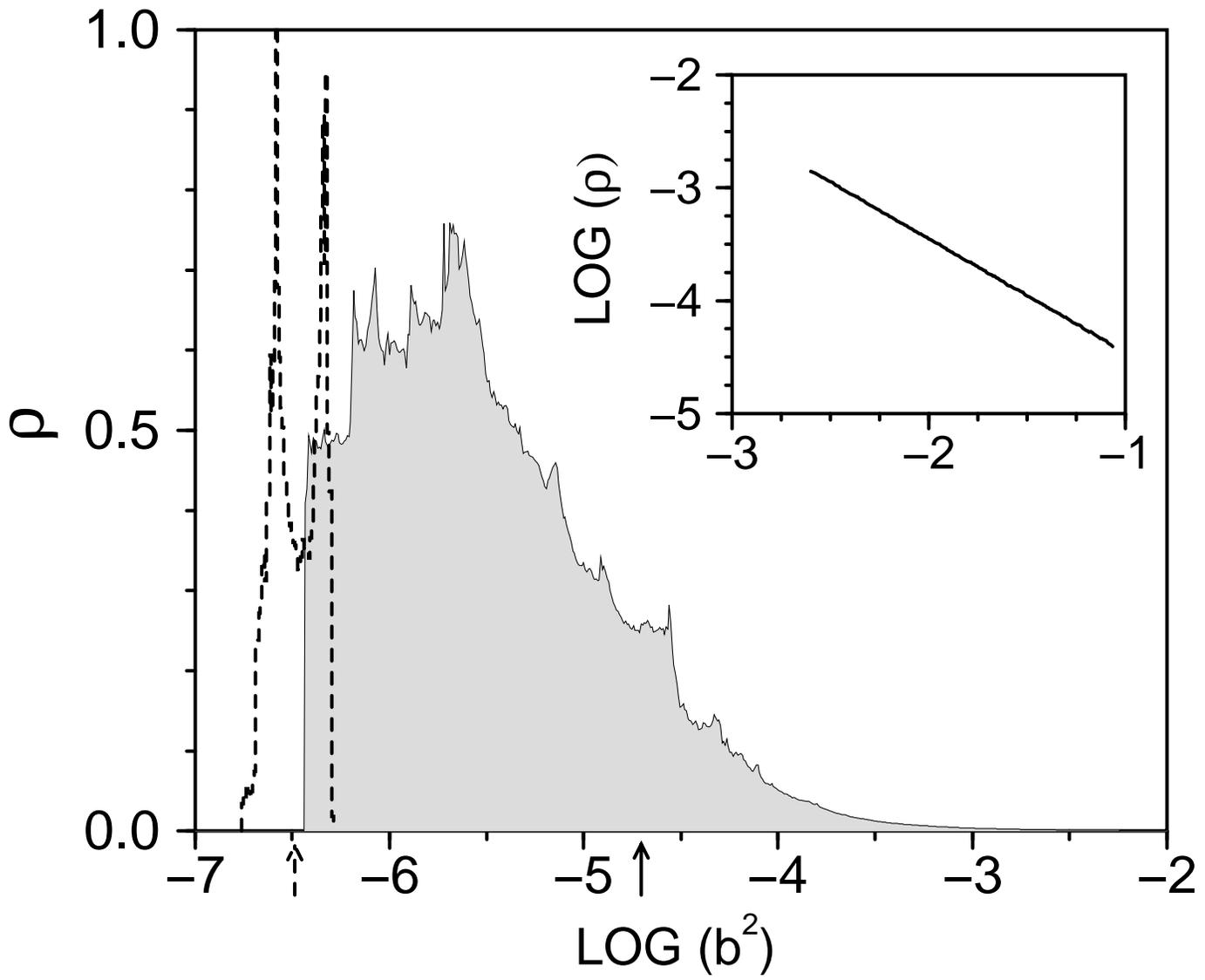
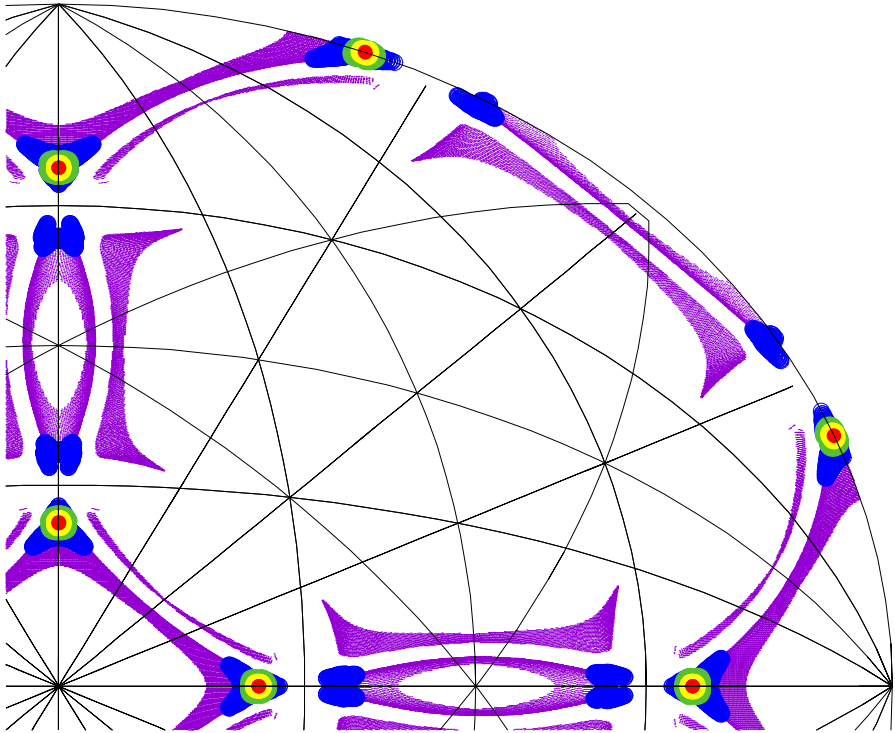
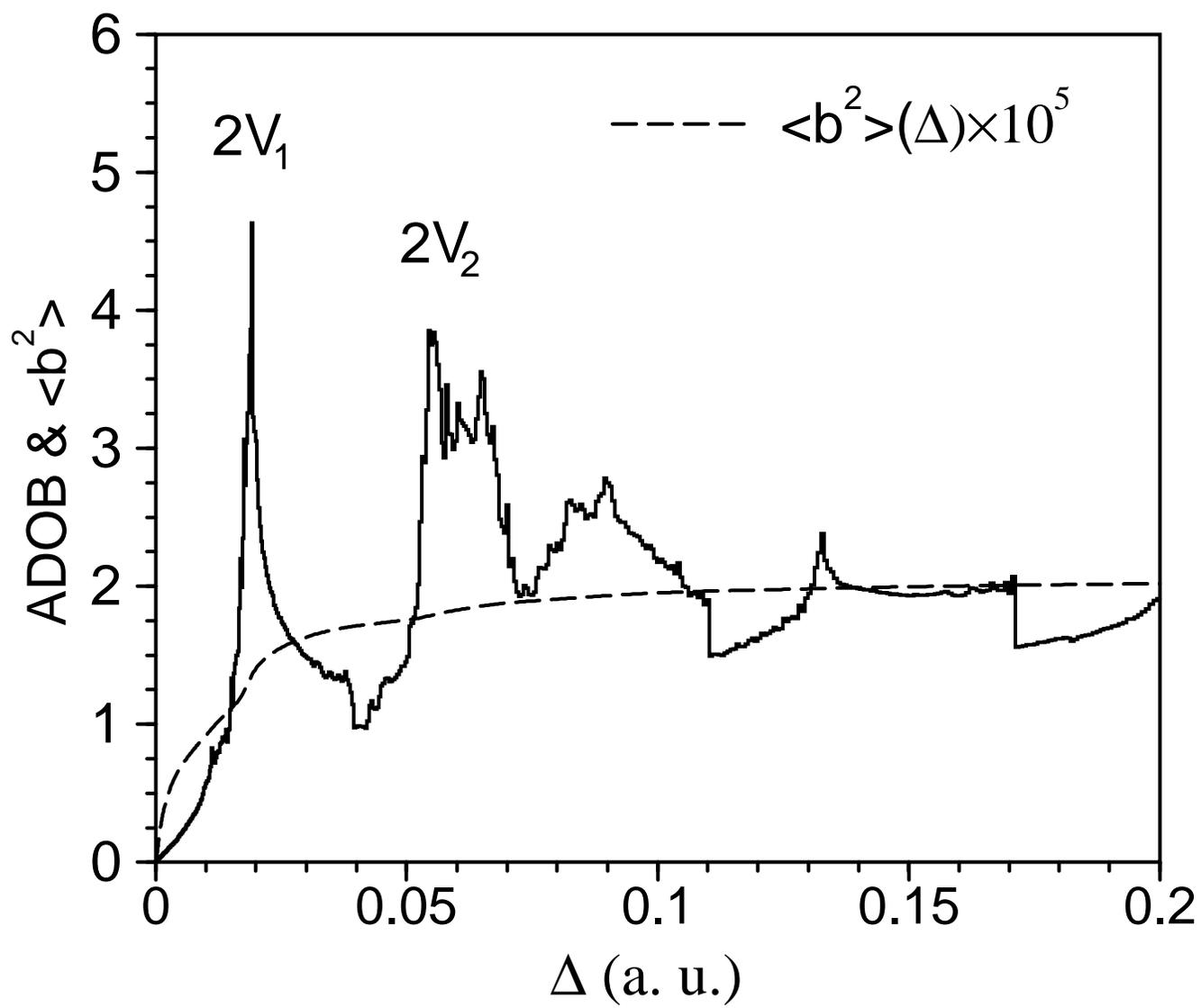
\end{document}